\documentclass[aps,pra,reprint,superscriptaddress]{revtex4-1}

\usepackage{graphicx}
\usepackage{bm}
\usepackage{bbold}
\usepackage{amssymb}
\usepackage{amsmath}
\newcommand{\red}[1]{#1}

\renewcommand{\vec}[1]{{\mathbf #1}}

\begin{document}

\title{Transport of light through a dense ensemble of cold atoms in a static electric field}

\author{S.E. Skipetrov}
\email[]{Sergey.Skipetrov@lpmmc.cnrs.fr}
\affiliation{Univ. Grenoble Alpes, CNRS, LPMMC, 38000 Grenoble, France}

\author{I.M. Sokolov}
\email[]{ims@is12093.spb.edu}
\affiliation{Department of Theoretical Physics, Peter the Great St. Petersburg Polytechnic University, 195251 St. Petersburg, Russia}

\date{\today}

\begin{abstract}
We demonstrate that the transport of coherent quasiresonant light through a dense cloud of immobile two-level atoms subjected to a static external electric field can be described by a simple diffusion process up to atomic number densities of the order of at least $10^2$ atoms per wavelength cubed. Transport mean free paths well below the wavelength of light in the free space can be reached without inducing any sign of Anderson localization of light or of any other mechanism of breakdown of diffusion.
\end{abstract}

\maketitle

\section{Introduction}
\label{sec:intro}
An ensemble of $N \gg 1$ identical, immobile two-level atoms randomly distributed in space with a given average number density $\rho$ represents a convenient theoretical model to study the multiple scattering of light \cite{lagen96,devries98}. On the one hand, such a physical system can be created experimentally by cooling an initially hot atomic vapor to sufficiently low temperatures using modern laser cooling techniques \cite{phillips98}. The predictions of the model can then be directly applied to describe experiments. On the other hand, two-level atoms are resonant point scatterers and as such can serve as a minimal model to study resonant scattering of light by more complex objects (small dielectric spheres or semiconductor grains, etc.). Surprisingly enough, even such a simplified model turns out to be difficult to treat analytically once the number of atoms per wavelength cubed of the resonant light $\lambda_0$ becomes significant \cite{nieuwen94,tiggelen94,cherroret16}. Numerical analysis of light scattering by resonant point scatterers has become a powerful tool to test analytic theories \cite{jenkins16,kwong18} and to explore fundamental phenomena \cite{fayard15,cottier18} in multiple scattering.

We have recently demonstrated that Anderson localization of light cannot be achieved in a three-dimensional (3D) ensemble of two-level atoms or, equivalently, point scatterers \cite{skip14} (a similar conclusion has been reached be Bellando \textit{et al.} \cite{bellando14}) unless the atoms are subjected to a strong magnetic field \cite{skip15}. Anderson localization is a wave interference phenomenon consisting in a halt of wave transport through a disordered system due to strong destructive interferences of scattered waves \cite{anderson58,lagendijk09,abrahams10}. It can take place for various types of waves, including `Schr\"{o}dinger waves' describing electrons in disordered solids \cite{anderson58,kramer93}, matter waves realized with cold atoms \cite{chabe08,jendr12}, sound \cite{hu08,cobus18}, or electromagnetic waves \cite{chabanov00,schwartz07}. In the latter case, however, no reliable experimental observation exists up to date in 3D \cite{skip16}. Our proposal of a 3D experiment with cold atoms in a static magnetic field \cite{skip16pra} requires a strong field that may be difficult to realize in practice. We have therefore explored the possibility of using an electric field instead, hoping that the Stark effect might have the same impact on localization as the Zeeman effect does \cite{skip19}. It turned out, however, that a static electric field does not induce Anderson localization of light in the atomic medium. A question then arises: what is the nature of optical transport in a dense ensemble of resonant atoms under conditions when Anderson localization takes place for scalar waves? Does the transport remain diffusive or does a new transport regime arise? In the present paper we provide answers to these question by computing the spatial distribution of the average excited state population in a dense 3D atomic system illuminated by a monochromatic plane wave that is quasi-resonant with one of the atomic transitions. We also compute the transmission coefficient of light through the atomic system. In both cases we find results that are perfectly compatible with the predictions of the diffusion theory in which an anisotropy of the atomic medium induced by the electric field is taken into account. This demonstrates that the transport of light remains diffusive even when the scattering is very strong and Anderson localization of light could be expected from naive arguments.

\section{The model}
\label{model}

We consider a cylindrical atomic sample of thickness $L$ and radius $R \gg L$,
\red{the cylinder axis coinciding with the $z$ axis of the coordinate system}
[see the inset of Fig.\ \ref{figlow}]. $N \gg 1$ identical immobile atoms are placed at randomly chosen, uncorrelated points $\left\{ \vec{r}_j \right\}$, $j = 1, \ldots, N$, inside the sample volume $V = \pi R^2 L$ with an average density $\rho = N/V$.
\red{In contrast to some other authors who model light scattering in atomic systems using similar approaches \cite{kwong18,araujo17}, we do not introduce an ``exclusion volume'' around each atom, so that two atoms has a nonvanishing probability of being arbitrary close to each other. Introducing an exclusion volume induces correlations between atomic positions---an additional complication that we wish to avoid here.}

\red{We assume that} each atom has a nondegenerate ground state $|{\cal E}_g, J_g = 0\rangle$ and three degenerate excited states $|{\cal E}_e, J_e = 1, m \rangle$, where ${\cal E}_{g,e}$ and $J_{g,e}$ are the energies and the total angular momenta of the ground and excited states, respectively, and $m = 0, \pm 1$ is the magnetic quantum number. The natural line width of the excited states is $\Gamma_0$. A spatially uniform, static external electric field $\vec{E}_{\mathrm{ext}}$ is applied to the system. Here we consider only $\vec{E}_{\mathrm{ext}}$ that is either parallel on perpendicular to the axis of the sample.
\red{The quantization axis is always parallel to $\vec{E}_{\mathrm{ext}}$. The field}
shifts the energies of the ground and excited states to new values ${\cal E}_g'$ and ${\cal E}_e'(m)$, respectively, due to the Stark effect \cite{friedrich90,sobelman92}. The degeneracy of the excited states is now partially lifted because ${\cal E}_e'(0) \ne {\cal E}_e'(-1) = {\cal E}_e'(1)$. The precise values of Stark shifts may depend on the magnitude of the applied field and on other parameters that are not included in our theoretical model, but important for us will be the energy differences ${\cal E}_e'(0) - {\cal E}_g' = \hbar \omega_0$ and ${\cal E}_e'(0) - {\cal E}_e'(\pm 1) = \hbar \Delta$. The Hamiltonian of such an atomic system interacting with the free electromagnetic field has been given previously \cite{skip19} and is reproduced by Eq.\ (\ref{ham}) of the Appendix \ref{app:qm}. It reduces to an effective non-Hermitian Hamiltonian $G$ given by Eq.\ (\ref{green}) and already discussed previously in Ref.\ \cite{skip19} where the eigenvalues and eigenfunctions of $G$ have been analyzed. The response of the atomic system to an external excitation can be expressed via the resolvent of the matrix $G$:
\begin{eqnarray}
{\cal R}(\omega) = \left[(\omega-\omega_0) \mathbb{1} + (\Gamma_0/2) G \right]^{-1},
\label{resolvent}
\end{eqnarray}
where $\mathbb{1}$ is a $3N \times 3N$ identity matrix.

From here on we assume that the atomic sample is illuminated by a monochromatic plane wave [probe light in the inset of Fig.\ \ref{figlow}] with a frequency $\omega$ and a wave vector $\vec{k_{\mathrm{in}}} = (\omega/c) \vec{e}_z$:
$\mathbf{E}_{\mathrm{in}}(\mathbf{r}) = \mathbf{u}_{\mathrm{in}} E_0 \exp(i \mathbf{k}_{\mathrm{in}} \mathbf{r})$, where the unit vector $\mathbf{u}_{\mathrm{in}}$ ($|\mathbf{u}_{\mathrm{in}}| = 1$) determines the polarization of the incident light.
Assume that $\Delta V$ is a small volume centered at $\vec{r}$. Then, the population of excited states corresponding to the magnetic quantum number $m$, per unit volume, is given by \cite{fofanov13}
\begin{eqnarray}
P_m(\vec{r}, \omega)  &=& \lim\limits_{\Delta V \to 0} \frac{1}{\hbar^2 \Delta V} \left| \sum\limits_{\vec{r}_j \in \Delta V}
\sum\limits_{n, m'} {\cal R}_{e_{jm} e_{nm'}} \right.
\nonumber \\
&\times& \left. \vec{d}_{e_{n m'} g_n} \cdot \vec{E}_{\mathrm{in}}(\vec{r}_n)
\vphantom{\sum\limits_{\vec{r}_j \in \Delta V}}
\right|^2,
\label{popm}
\end{eqnarray}
where $\vec{d}_{e_{n m'} g_n} = \langle {\cal E}_e'(m'), J_{e}=1, m'|{\hat{\mathbf{D}}}_n | {\cal E}_g', J_{g} = 0 \rangle$ is the matrix element of the dipole moment operator.

The translational symmetry imposes that for an infinitely wide sample ($R \to \infty$), the average value of $P_m(\vec{r}, \omega)$ would be a function of $z$ only: $\langle P_m(\vec{r}, \omega) \rangle = \langle P_m(z, \omega) \rangle$. Here the angular brackets $\langle \cdots \rangle$ denote averaging over different atomic configurations $\left\{ \vec{r}_j \right\}$. Obviously, in a cylindrical sample of finite radius $R$, $\langle P_m(\vec{r}, \omega) \rangle$ keeps a dependence on the transverse position $\vec{r}_{\perp} = \left\{ x, y \right\}$. However, for $R \gg L$ we can still obtain a meaningfull quantity that depends only on $z$ and approximates $\langle P_m(z, \omega) \rangle$ of an infinitely wide sample by averaging over the central part of our cylindrical sample:
\begin{eqnarray}
\langle P_m(z, \omega) \rangle = \frac{1}{\pi R_1^2} \int\limits_{r_{\perp} < R_1 < R} \langle P_m(\vec{r} = \left\{ \vec{r}_{\perp}, z \right\}, \omega) \rangle d^2 \vec{r}_{\perp}.\;\;\;\;\;
\label{popm1}
\end{eqnarray}

Finally, we will be interested in the total population of all the three possible excited states:
\begin{eqnarray}
\langle P(z, \omega) \rangle = \sum\limits_{m=-1}^1 \langle P_m(z, \omega) \rangle.
\label{popsum}
\end{eqnarray}

Strictly speaking, $\langle P(z, \omega) \rangle$ is not equal to the average intensity of light in the system $\langle I(z, \omega) \rangle$.
\red{However, a linear relation between $\langle P(z, \omega) \rangle$ and $\langle I(z, \omega) \rangle$ turns out to be a good approximation \cite{lax51}, in particular, in a dilute medium where the diffuse behavior of $\langle I(z, \omega) \rangle$ implies the diffuse behavior of $\langle P(z, \omega) \rangle$.}
In the next section, we will compare our results for $\langle P(z, \omega) \rangle$ with predictions of a simple diffusion theory.
We will analyze $\langle P(z, \omega) \rangle$ as a function of frequency $\omega$  and angle between $\vec{k}_{\mathrm{in}}$ and $\vec{E}_{\mathrm{ext}}$ by numerically evaluating Eq.\ (\ref{popm}), averaging over many different atomic configurations $\{ \vec{r}_j \}$ and over the central part of the considered cylindrical sample according to Eq.\ (\ref{popm1}), and then summing over $m$ as defined by Eq.\ (\ref{popsum}).

\section{Average population of excited states}
\label{population}

In our system, the incident light that is quasiresonant with the transition $|{\cal E}_g', J_g = 0 \rangle \to |{\cal E}_e'(0), J_e = 1, m = 0 \rangle$  (i.e., $\omega \simeq \omega_0$), will be most efficiently scattered if it is linearly polarized along the external electric field $\vec{E}_{\mathrm{ext}}$  which, in its turn, is perpendicular to the incident wave vector $\vec{k}_{\mathrm{in}}$. We will symbolically denote such a linear polarization by $\vec{u}_{\mathrm{in}} = \;\uparrow$. Combining  $\vec{E}_{\mathrm{ext}} \perp \vec{k}_{\mathrm{in}}$ with $\omega \simeq \omega_0$ and $\vec{u}_{\mathrm{in}} = \;\uparrow$ ensures the strongest scattering starting from first scattering event and hence the fastest realization of the multiple scattering regime. On the other hand, when $\vec{E}_{\mathrm{ext}} \parallel \vec{k}_{\mathrm{in}}$, the strongest scattering is reached for the light that is quasiresonant with one of the transitions  $|{\cal E}_g', J_g = 0 \rangle \to |{\cal E}_e'(\pm1), J_e = 1, m = \pm 1 \rangle$  (i.e., $\omega \simeq \omega_0 - \Delta$) and circularly polarized. We will denote such a polarization by $\vec{u}_{\mathrm{in}} = \;\circlearrowleft$. For convenience of comparing results corresponding to the two aforementioned combinations of frequencies and polarizations, we will count the frequency detuning $\delta$ from the resonant frequency of the corresponding transition. Thus, in the rest of the paper, the same value $\delta$ will correspond to $\omega - \omega_0 = \delta$ for $\vec{E}_{\mathrm{ext}} \perp \vec{k}_{\mathrm{in}}$, $\vec{u}_{\mathrm{in}} = \;\uparrow$ but to $\omega - (\omega_0 - \Delta) = \delta$ for $\vec{E}_{\mathrm{ext}} \parallel \vec{k}_{\mathrm{in}}$, $\vec{u}_{\mathrm{in}} = \;\circlearrowleft$ because the resonance frequency of the probed transition is $\omega_0 - \Delta$ in the latter case. Needless to say that in the absence of external fields (i.e., for $\Delta = 0$), the average excited state population $\langle P(z, \omega) \rangle$  is independent of the polarization of incident light.

\begin{figure}
\includegraphics[width=0.95\columnwidth]{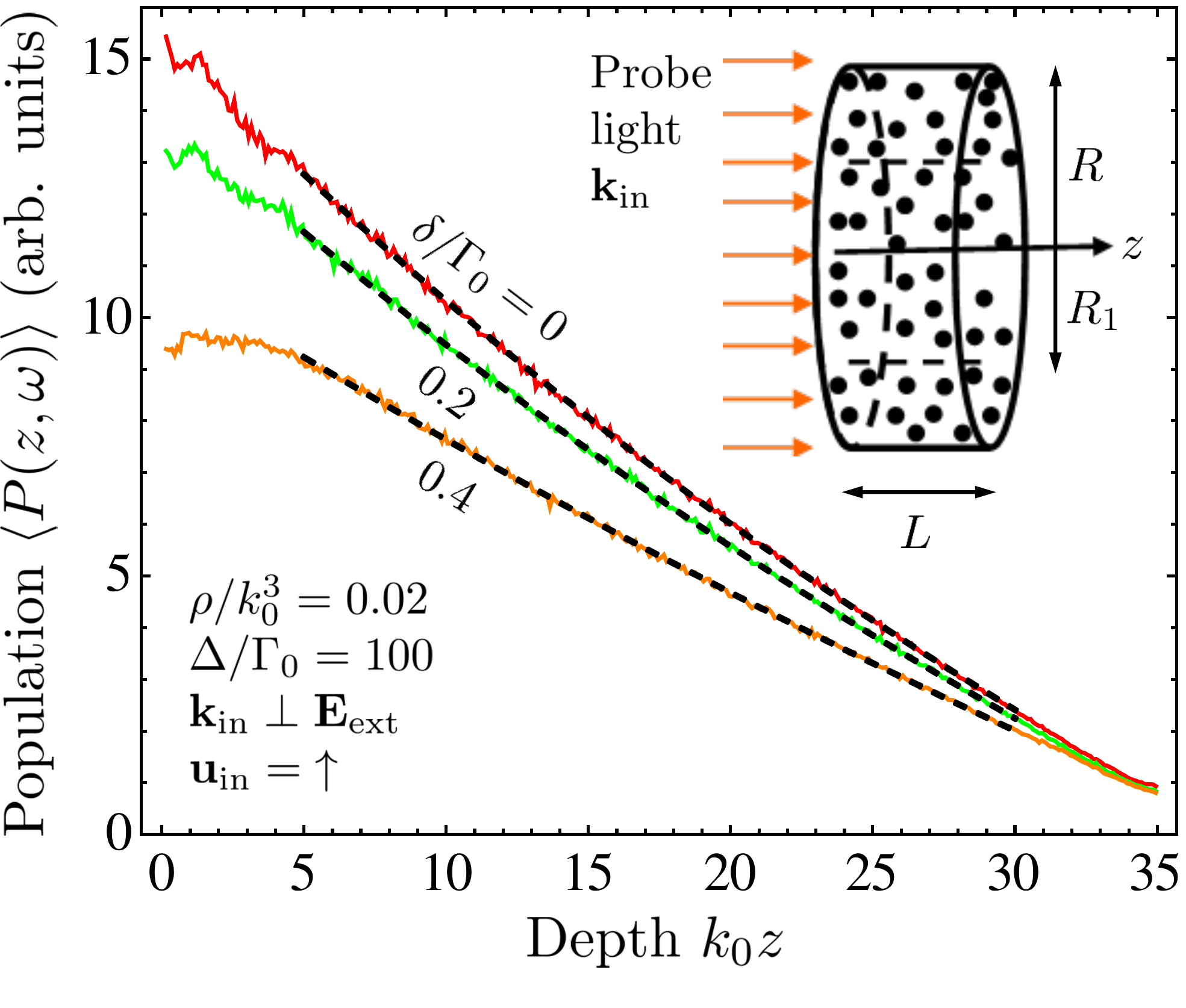}
\caption{Average population of excited states $\langle P(z, \omega) \rangle$ for different frequencies of a linearly polarized incident wave and an external electric field perpendicular to its direction of propagation, for a dilute atomic medium. Averaging is performed over 13000 independent atomic configurations for each curve. The inset illustrates the considered experimental geometry; $k_0 L = 35$, $k_0 R = 70$ and $k_0 R_1 = 35$ for this figure. Diffusion-theory fits (dashed lines) are performed for the data corresponding to $k_0 z \in [5, 30]$ using Eq.\ (\ref{sol3}).}
\label{figlow}
\end{figure}

We first consider a relatively dilute medium in which light transport is expected to be diffusive \cite{ishimaru78,sheng95,akkermans07}. Examples of spatial profiles of $\langle P(z, \omega) \rangle$ obtained for the linear polarization of the incident wave and different detunings $\delta$ are shown in Fig.\ \ref{figlow}. Whereas diffusion theory \cite{ishimaru78,sheng95,akkermans07} predicts a simple linear decay of $\langle P(z, \omega) \rangle$ far from the boundaries of an infinitely wide slab and for $R \to \infty$ [see Eq.\ (\ref{sol2a}) in the Appendix \ref{app:dif}], the curves of Fig.\ \ref{figlow} exhibit weak but visible concavity. It can be explained by taking into account the finite radius $R$ of the considered cylindrical sample and the anisotropy of light transport induced by the external electric field. We start with an anisotropic photon diffusion equation for the
\red{average population of excited states $\red{P_{\mathrm{dif}}}(\vec{r})$} inside the cylindrical sample depicted in the inset of Fig.\ \ref{figlow}:
\begin{eqnarray}
-\bm{\nabla}_{\vec{r}} \cdot {\tilde D} \cdot \bm{\nabla}_{\vec{r}} \red{P_{\mathrm{dif}}}(\vec{r}) = P_0 \ell_z^* \delta(z-\ell_{z}^*)
\Pi\left(\frac{r_{\perp}}{2R} \right),
\label{difeq}
\end{eqnarray}
where the right-hand side describes the source of diffuse waves due to the coherent incident plane wave of intensity $I_0 \red{\propto P_0}$ that is assumed to be converted into a diffuse one at a distance $\ell_z^*$ from the front surface of the sample. The factor $\ell_z^* \delta(z-\ell_{z}^*)$ on the right-hand side replaces a more accurate source function $\exp(-z/\ell_z^*)$ describing the progressive isotropization of the incident radiation as it enters into the disordered medium. $\Pi(x)$ is the normalized boxcar function [$\Pi(x) = 1$ for $|x| < 1/2$, $\Pi(x) = 0$ for $|x| > 1/2$] describing the fact that the conversion of the incident coherent light into diffuse radiation takes place only inside the sample (i.e., for $r_{\perp}< R$). The diffusion tensor is $D = {\tilde D}/\tau^*$ with
\begin{eqnarray}
{\tilde D} = \frac{1}{3} \left(
\begin{matrix}
\ell_{\perp}^{*2} & 0 & 0 \\
0 & \ell_{\perp}^{*2} & 0 \\
0 & 0 & \ell_z^{*2}
\end{matrix}
\right)
\label{tensord}
\end{eqnarray}
and $\tau^*$ the transport mean free time. The transport mean free paths $\ell_{\perp}^*$ and $\ell_{z}^*$ in the $xy$ plane and along the $z$ axis, respectively, can be different because of the external electric field that breaks the rotational symmetry and makes the atomic system anisotropic.
\red{Strictly speaking, Eq.\ (\ref{tensord}) with $\ell_z^* \ne \ell_{\perp}^* = \ell_x^* = \ell_y^*$ holds when $\vec{E}_{\mathrm{ext}} \parallel \vec{k}_{\mathrm{in}} \parallel \vec{e}_z$ and the axes $x$ and $y$ are both perpendicular to the external field.
When $\vec{E}_{\mathrm{ext}} \perp \vec{k}_{\mathrm{in}}$ (say, $\vec{E}_{\mathrm{ext}} \parallel \vec{e}_x$), we will have $\ell_x^* \ne \ell_y^* = \ell_z^*$ and $\ell_{\perp}^*$ cannot be introduced. However, we will study only quantities that are integrated over a sufficiently large area in the $xy$ plane [see, e.g., Eq.\ (\ref{popm1})] and thus accounting for $\ell_x^* \ne \ell_y^*$ will not be essential for us. We will instead use an effective value $\ell_{\perp}^*$ for both $\ell_x^*$ and $\ell_y^*$ even when $\vec{E}_{\mathrm{ext}} \perp \vec{k}_{\mathrm{in}}$.       }

Anisotropic diffusion of light has been previously studied both theoretically \cite{stark96,tiggelen96} and experimentally \cite{kao96,wiersma99,johnson08}. Our Eq.\ (\ref{tensord}) takes into account the fact that the transport mean free time $\tau^*$ in an atomic medium is mainly determined by the lifetime $1/\Gamma_0$ of the atomic excited states: $\tau^* \simeq 1/\Gamma_0$ \cite{labeyrie03}, and hence $D = \ell^{*2}/3 \tau^*$. We supplement Eq.\ (\ref{difeq}) by boundary conditions \cite{ishimaru78,akkermans07,zhu91}:
\begin{eqnarray}
\label{bc1}
&& \red{P_{\mathrm{dif}}}(\vec{r} = \{\vec{r}_{\perp}, z \}) = 0,\;\; z = -h_z,\, L + h_z,
\\
\label{bc2}
&& \red{P_{\mathrm{dif}}}(\vec{r} = \{\vec{r}_{\perp}, z \}) = 0,\;\; r_{\perp} = R + h_{\perp},
\end{eqnarray}
where $h_z$ and $h_{\perp}$ are the so-called extrapolation lengths in the longitudinal and transverse directions. They account for internal reflections of multiply scattered waves at the boundaries of the disordered sample and are typically of the order of corresponding transport mean free paths: $h_z \sim \ell_z^*$, $h_{\perp} \sim \ell_{\perp}^*$  \cite{zhu91}.

The solution of the anisotropic photon diffusion equation (\ref{difeq}) for the geometry corresponding to our numerical calculations is presented in Appendix \ref{app:dif}. The resulting Eq.\ (\ref{sol3}) for $\red{P_{\mathrm{dif}}}$ provides very good fits to our numerical data in the central part of the sample, see dashed lines in Fig.\ \ref{figlow}. The free parameters of the fits are $\ell_z^*$, $\ell_{\perp}^*$, $h_z$, $h_{\perp}$ and a constant prefactor $C$ in front of Eq.\ (\ref{sol3}) to adjust the overall magnitude of intensity. The best-fit parameters fall in reasonable ranges: $k_0 \ell_z^* = 1.4$--1.7, $k_0 \ell_{\perp}^* = 1.5$--2.4, $h_z/\ell_z^* \sim 1.4$--2.2 and $h_{\perp}/\ell_{\perp}^* \sim 0.8$--1.3. In particular, the transport mean free paths are comparable with the value $\ell_0$ expected on resonance ($\omega = \omega_0$) in the absence of external fields based on the perturbation theory in $\rho/k_0^3 \ll 1$: $k_0 \ell_0 = k_0^3/6\pi\rho \simeq 2.65$ for $\rho/k_0^3 = 0.02$ in Fig.\ \ref{figlow}. However, it may be dangerous to consider these values as reliable estimates of real physical parameters because of the large number (five) of free parameters in our fits. Other combinations of $\ell_z^*$, $\ell_{\perp}^*$, $h_z$, $h_{\perp}$ and $C$ may provide fits of comparable quality.

\begin{figure*}
\includegraphics[width=0.95\columnwidth]{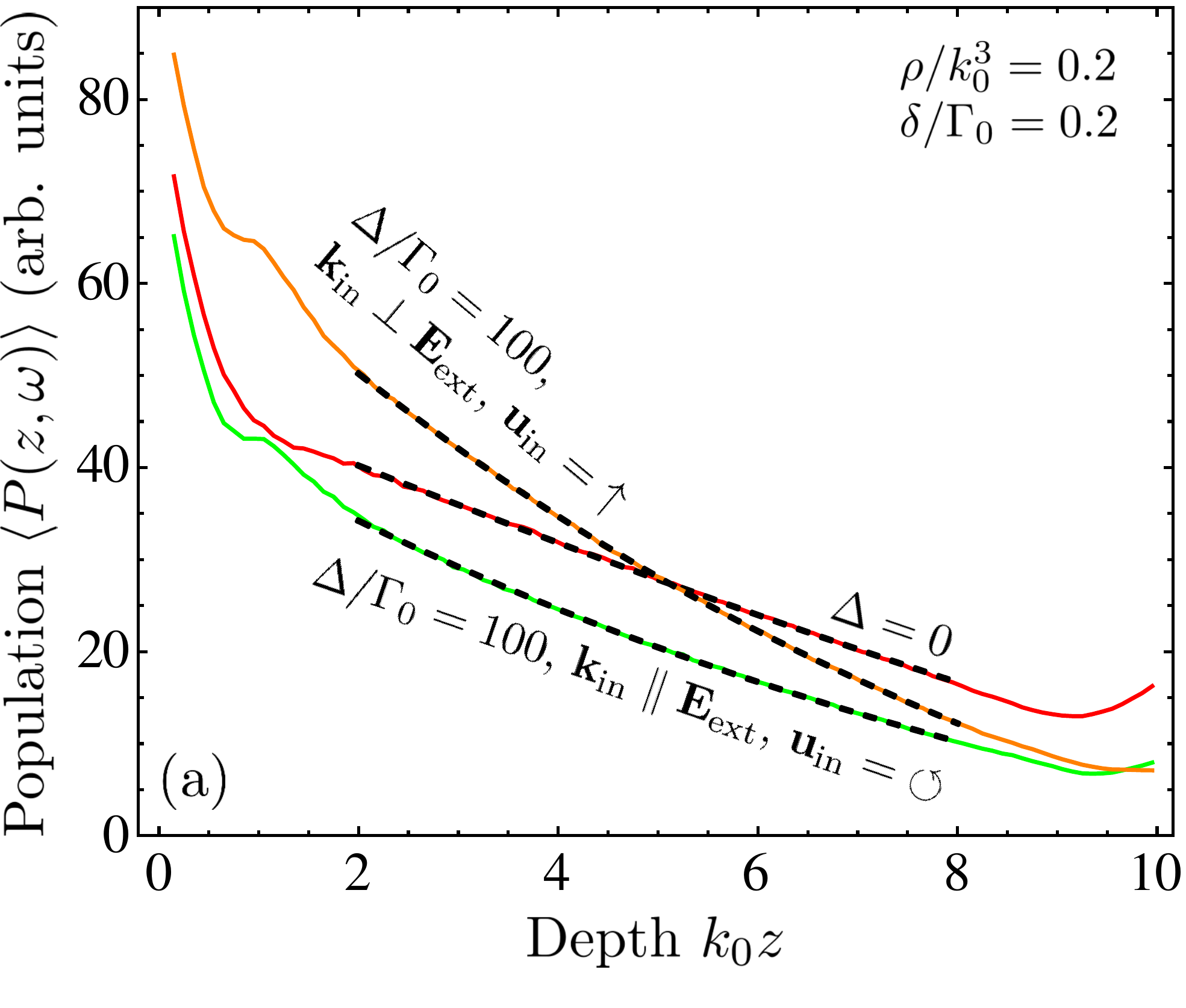} \hspace{5mm}
\includegraphics[width=0.95\columnwidth]{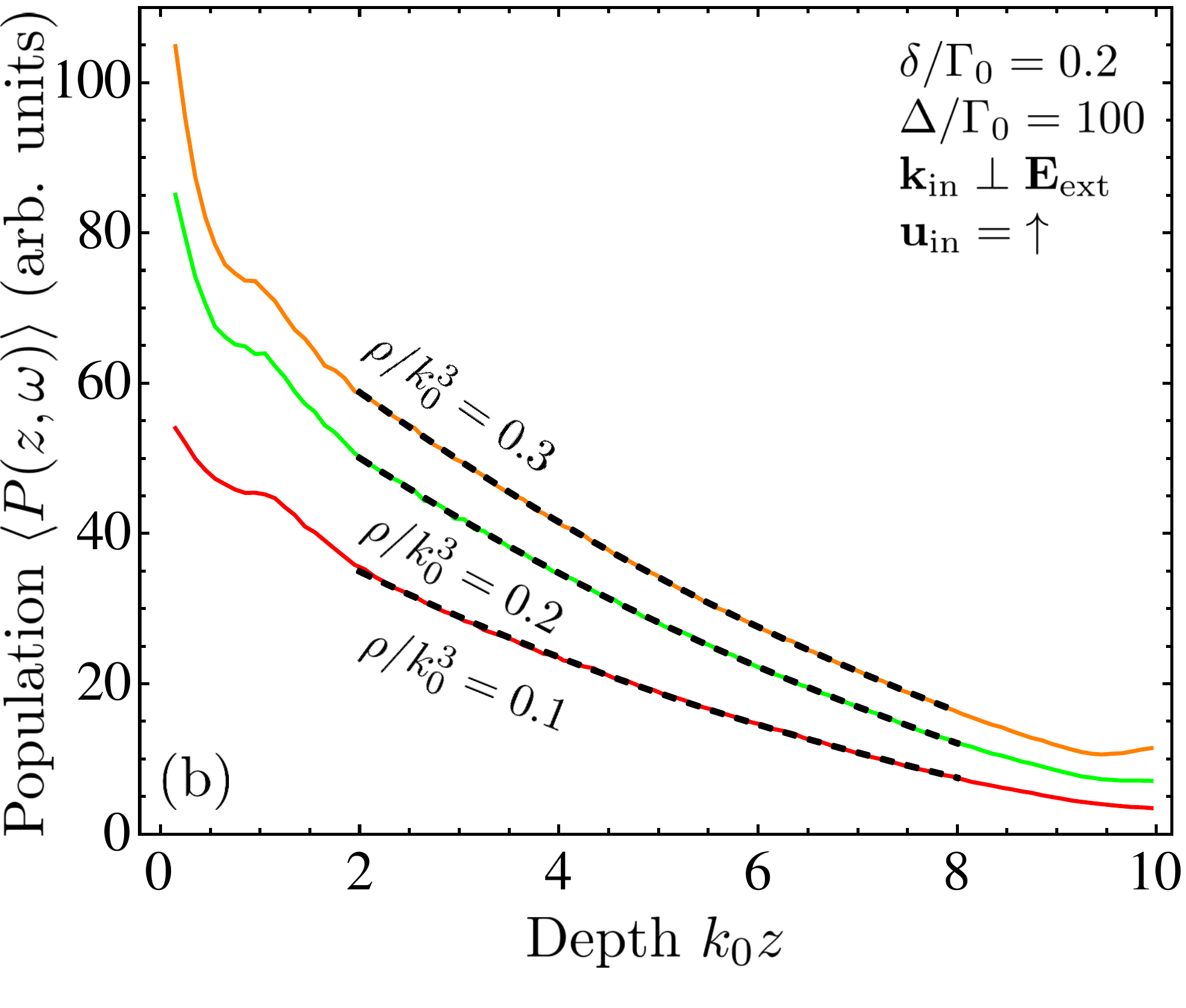}
\includegraphics[width=0.95\columnwidth]{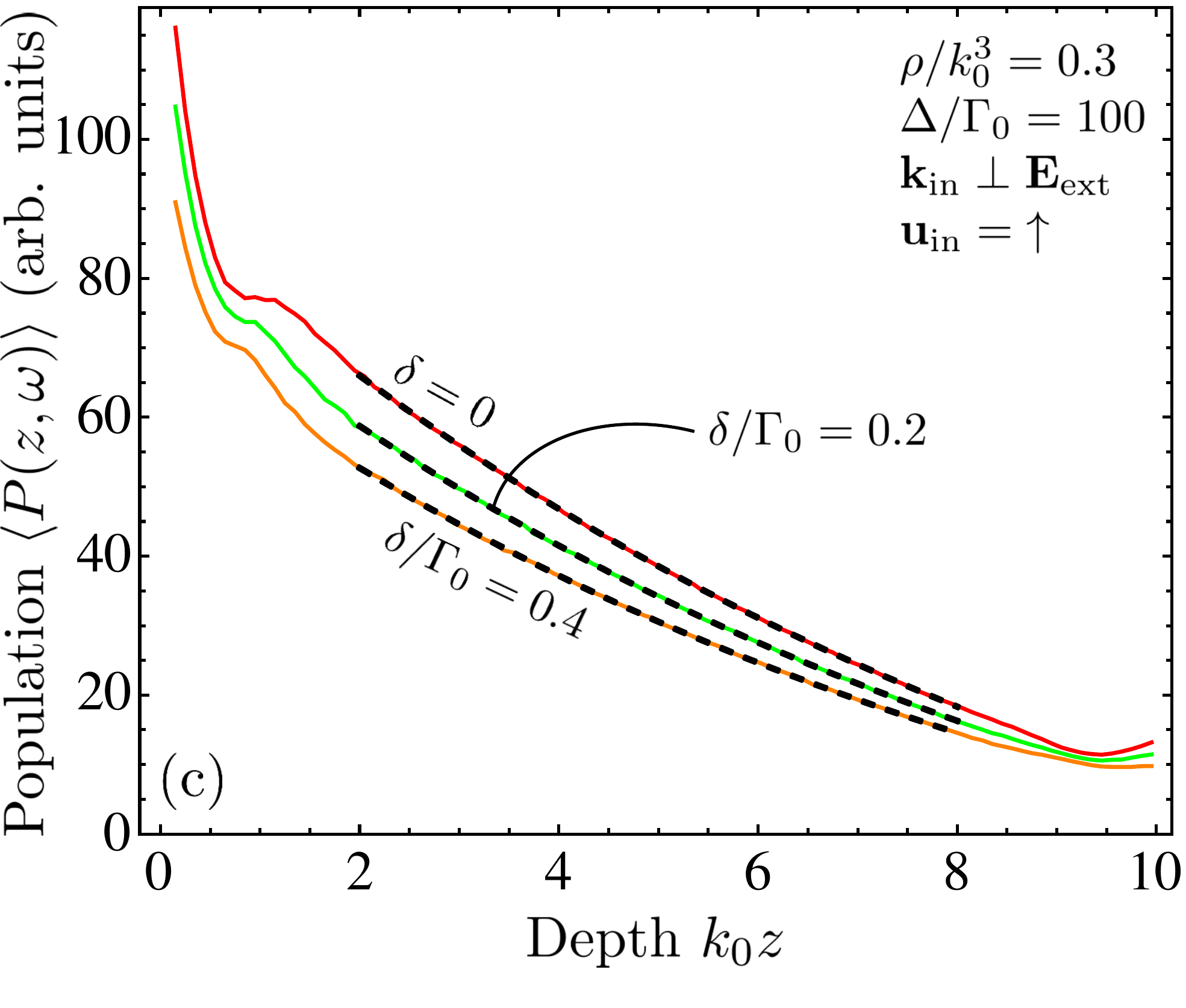} \hspace{5mm}
\includegraphics[width=0.95\columnwidth]{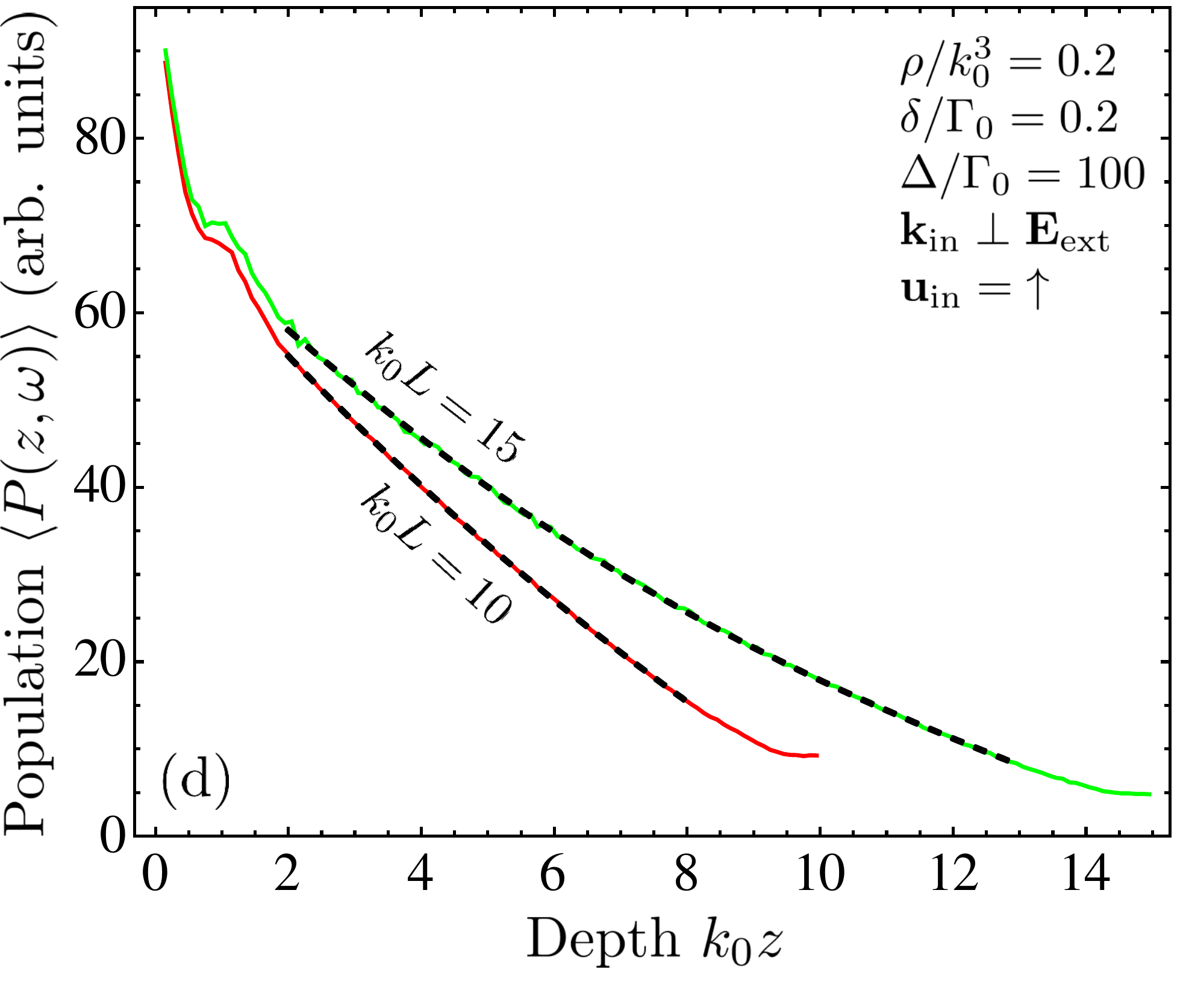}
\caption{Average population of excited states $\langle P(z, \omega) \rangle$ for different frequencies and polarizations of the incident wave and for a dense atomic medium.
(a) Comparison of results in the absence of external fields (red curve, $\Delta = 0$) with those in a strong electric field (the green and orange curves, $\Delta/\Gamma_0 = 100$).
(b) Comparison of results obtained at different atomic number densities $\rho$.
(c) Comparison of results obtained at different detunings $\delta$.
(d) Comparison of results obtained for two different sample thicknesses $L$.
$k_0 L = 10$, $k_0 R = 20$, and $k_0 R_1 = 10$ for the panels (a)--(c); $k_0 R = 25$, $k_0 R_1 = 12$ for the panel (d).
Averaging is performed over at least 40000 independent atomic configurations for each curve.
Dashed lines in all panels show diffusion-theory fits [Eq.\ (\ref{sol3})] to the numerical results for $k_0 z \in [2, k_0 L - 2]$.
$\ell_z^* = \ell_{\perp}^*$ and $h_z = h_{\perp}$ were imposed for the fit to the data corresponding to $\Delta = 0$ in the panel (a) because the medium is isotropic in the absence of external fields.
}
\label{figpop}
\end{figure*}

\begin{figure*}
\includegraphics[width=0.98\columnwidth]{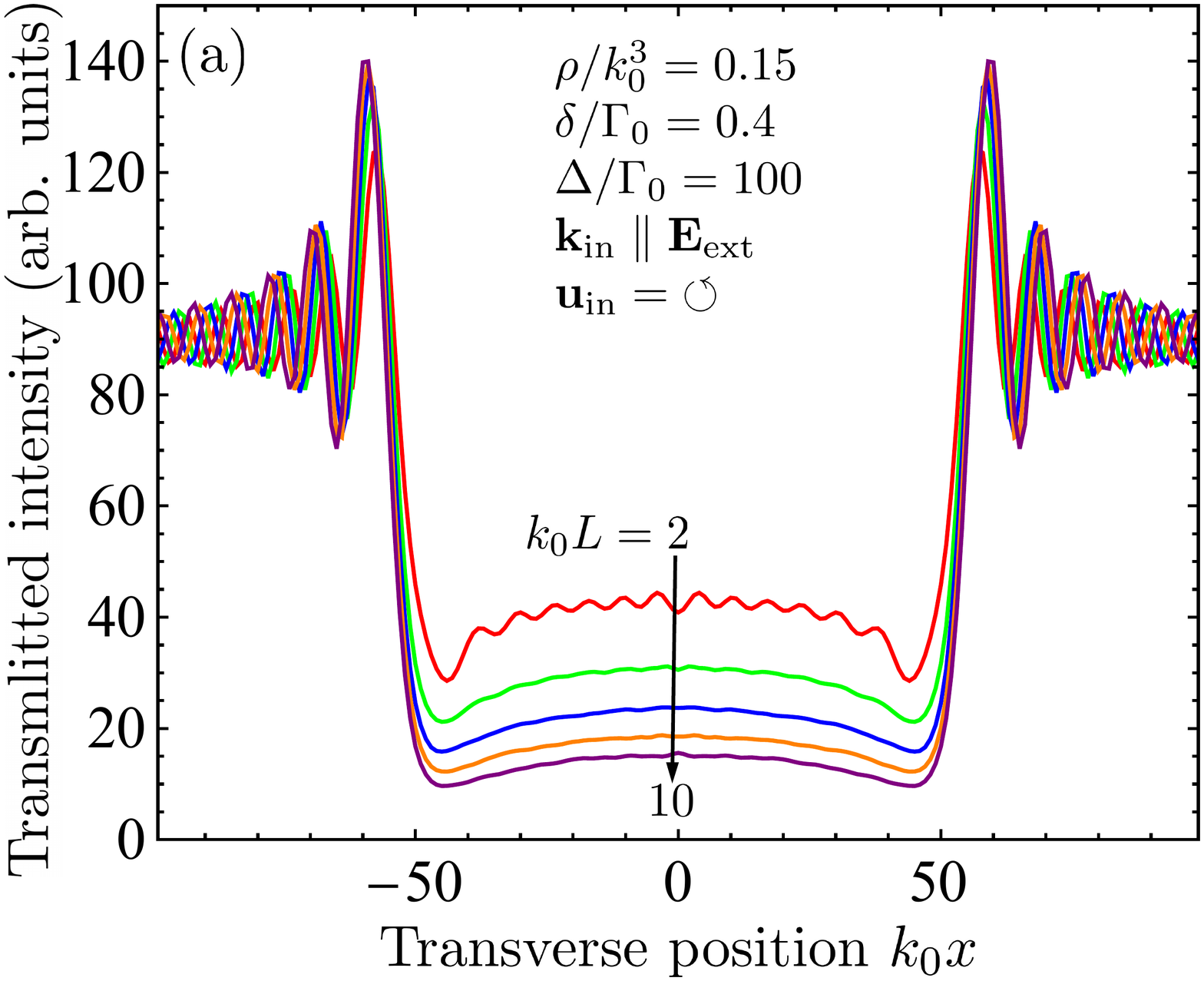} \hspace{-1.3cm}
\includegraphics[width=0.98\columnwidth]{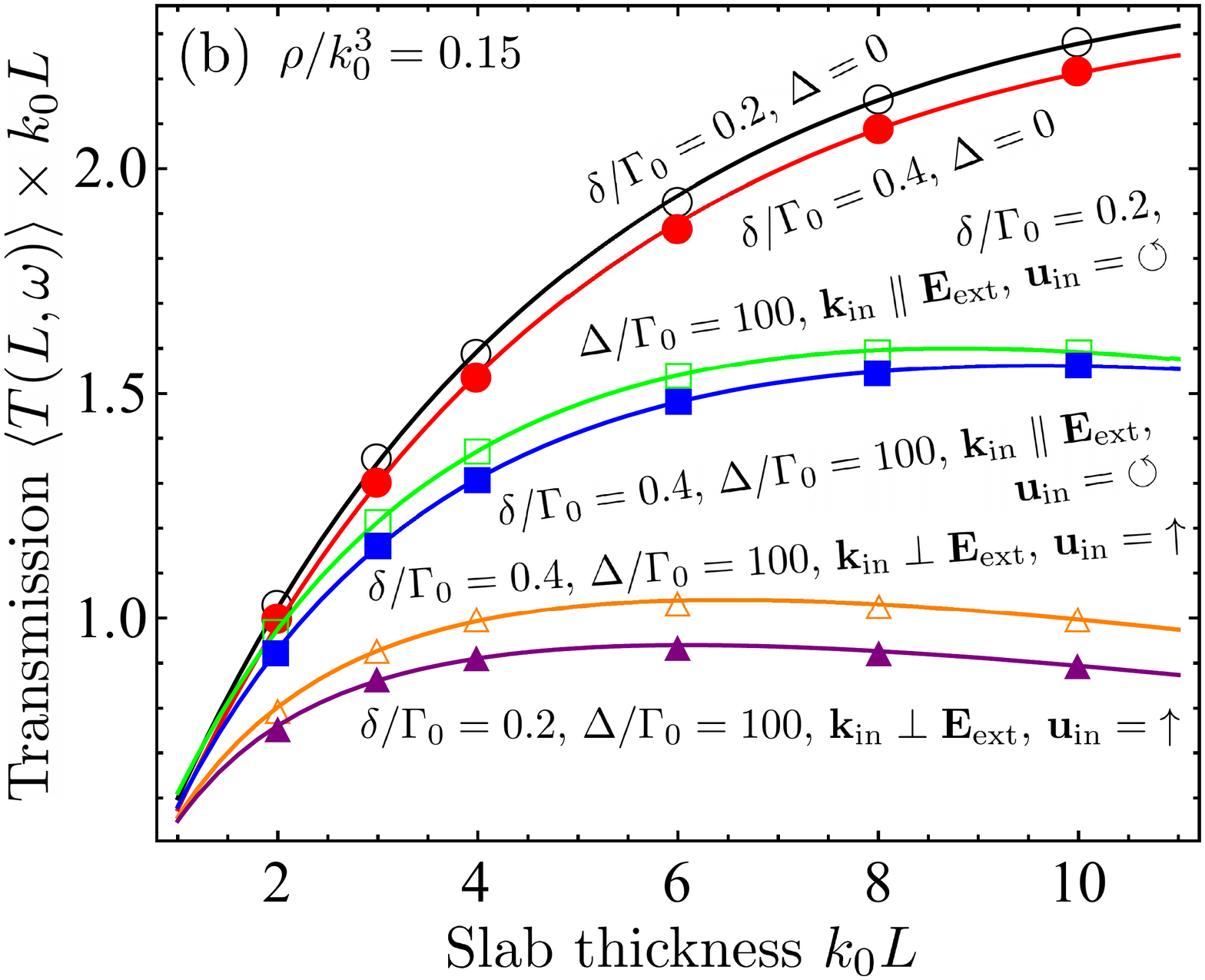}
\caption{(a) Average intensity of light transmitted through cylindrical atomic samples of radius $k_0 R = 50$ and different thicknesses $k_0 L = 2$--10, at a distance $k_0(z - L) = 10$ from the sample
and as a function of transverse position $x$ for $y = 0$, in a strong external electric field $\vec{E}_{\mathrm{ext}}$. The intensity shown in the panel (a) is averaged over $k_0 r_{\perp} < 35$ to obtain $\langle T \rangle$ in the panel (b) where symbols show the average transmission coefficient $\langle T \rangle$ of the cylindrical atomic sample multiplied by its thickness $k_0 L$ for different frequency detunings $\delta$ and different polarizations of the incident plane wave. Results obtained in the absence of the field ($\Delta = 0$) are compared with those in a strong electric field ($\Delta = 100$). Solid lines show diffusion-theory fits [Eq.\ (\ref{trans}) with $R_1 = R$] to numerical data. The best-fit parameters are given in Table\ \ref{tabfit}.}
\label{figtrans}
\end{figure*}

We now turn to dense atomic media where a breakdown of diffuse transport may be expected. Figure\ \ref{figpop}(a) shows that the decay of $\langle P(z, \omega) \rangle$ with depth $z$ inside the atomic sample still remains roughly linear for different polarizations of incident light, with or without the external electric field. Equation\ (\ref{sol3}) resulting from the diffusion theory provides excellent fits to the numerical data (dashed lines) similarly to the low-density case. The quality of fits remains very good for data corresponding to different densities of the atomic system [Fig.\ \ref{figpop}(b)], different frequencies of the incident light [Fig.\ \ref{figpop}(c)], and different thicknesses $L$ of the atomic sample [Fig.\ \ref{figpop}(d)]. This establishes the validity of diffusion theory for light transport in dense clouds of cold atoms in strong electric fields \red{at least} up to densities of the order of $10^2$ atoms per wavelength cubed.
\red{Even though we do not study densities larger than $\rho/k_0^3 = 0.3$ (which corresponds to $\rho \lambda_0^3 \simeq 75$) in this work, we expect this conclusion to hold at higher densities as well because no signatures of Anderson localization were found from the analysis of quasi-modes of dense atomic clouds up to $\rho/k_0^3 = 1.5$ \cite{skip19}. We expect scattering to weaken and homogenization to take place at even higher densities for which the atomic system should start to behave as a homogeneous medium with some effective properties. The effective optical properties of large atomic ensembles is a subject of intense current research \cite{jenkins16,morice95,java17}.
}

It is worthwhile to note that despite the demonstration of the validity of diffusion theory, we are not able to provide an analytic theory for the transport mean free paths $\ell_z^*$ and $\ell_{\perp}^*$ for dense scattering media where perturbation theory in $\rho/k_0^3 \ll 1$ fails. Calculation of transport properties of strongly scattering media remains a complicated theoretical problem even in the absence of external fields \cite{tiggelen94,cherroret16}.

\section{Average transmission}
\label{transmission}

Even if the spatial dependencies of the average population of excited states or, equivalently, of the average diffuse intensity inside a disordered atomic sample provide very useful information about the optical transport inside the sample, they are difficult to access experimentally. In a typical optical experiment, one measures the intensity $I$ of light transmitted through a disordered sample and having polarization $\vec{u}$ ($|\vec{u}| = 1$), or the transmission coefficient $T$. The intensity of light transmitted through the atomic sample can be written as a result of interference of incident and scattered waves:
\begin{eqnarray}
I(\vec{r}, \vec{u}, \omega) &=& \frac{c}{4\pi}
\left|
\vec{u}^* \cdot \vec{E}_{\mathrm{in}}(\vec{r}) \vphantom{\sum}
\right.
\nonumber \\
&+&  \left.
\frac{k^3}{\hbar} \sum\limits_{j,m} \sum\limits_{n,m'}
f_{e_{jm}}(\vec{r}, \vec{u}, \omega) {\cal R}_{e_{jm} e_{nm'}}(\omega)
\right.
\nonumber \\
&\times& \left.
\vec{d}_{e_{n m'} g_n} \cdot \vec{E}_{\mathrm{in}}(\vec{r}_n)
\vphantom{\sum}
\right|^{2},
\label{intensity}
\end{eqnarray}
where
\begin{eqnarray}
f_{e_{jm}}(\vec{r}, \vec{u}, \omega) &=&
\frac{e^{i k |\vec{r} - \vec{r}_j|}}{k |\vec{r} - \vec{r}_j|}
\left[ \vec{u}^* \cdot \vec{d}_{g_j e_{jm}}
\vphantom{\frac{[\vec{u}^* \cdot (\vec{r} - \vec{r}_j)]
[ \vec{d}_{g_j e_{jm}} \cdot (\vec{r} - \vec{r}_j)]}{|\vec{r} - \vec{r}_j|^2}}
\right.
\nonumber \\
&-& \left. \frac{[\vec{u}^* \cdot (\vec{r} - \vec{r}_j)]
[ \vec{d}_{g_j e_{jm}} \cdot (\vec{r} - \vec{r}_j)]}{|\vec{r} - \vec{r}_j|^2}
\right]\;\;\;\;
\label{fprop}
\end{eqnarray}
describes the propagation of light from the atom $j$ to a point $\vec{r}$, and $k = \omega/c$.

We calculate the average intensity of transmitted light that would be detected by a polarization-insensitive photodetector by summing over all directions of $\vec{u}$:
\begin{eqnarray}
\langle I(\vec{r},  \omega) \rangle &=&
\int\limits_{4\pi} d^2\vec{u} \langle I(\vec{r},  \vec{u}, \omega) \rangle.
\label{intensity2}
\end{eqnarray}
The average intensity $\langle I(\vec{r} = \{\vec{r}_{\perp},z \}, \omega) \rangle$ at a distance $k_0(z-L) = 10$ from the sample is shown in Fig.\ \ref{figtrans}(a) as a function of transverse position $x$ for $y = 0$ (remember that $\vec{r}_{\perp} = \{x, y \}$). Figure \ref{figtrans}(a) shows typical results for given atomic density $\rho/k_0^3 = 0.15$ and frequency detuning $\delta/\Gamma_0 = 0.4$ but similar results are obtained for other values of $\rho$ and $\delta$. The spatial profile of intensity exhibits an oscillatory diffraction pattern due to the finite extent of the sample in the transverse directions $x$, $y$ (i.e., due to the fact that $R < \infty$). The depth of the intensity drop near the sample axis (i.e., around $r_{\perp} = 0$) provides information about the average intensity transmission coefficient $\langle T \rangle$:
\begin{eqnarray}
\langle T(L, \omega) \rangle  = \frac{1}{I_1 \pi R_1^2} \int\limits_{r_{\perp} < R_1 < R} \langle I(\vec{r} = \left\{ \vec{r}_{\perp}, z \right\}, \omega) \rangle d^2 \vec{r}_{\perp},\;\;\;\;\;\;\;
\label{trans2}
\end{eqnarray}
where $I_1$ is the intensity obtained from Eqs.\ (\ref{intensity}) and (\ref{intensity2}) in the absence of the atomic sample.

\begin{table*}[t]
\caption{\label{tabfit} Best fit parameters for curves in Fig.\ \ref{figtrans}(b). The frequency difference $\Delta$ and the detuning $\delta$ are in units of the natural line width $\Gamma_0$. $\ell_z^* = \ell_{\perp}^*$ and $h_z = h_{\perp}$ were imposed at $\Delta = 0$ because the medium is isotropic in the absence of external fields.
\red{The values of the scattering mean free path $\ell_z$ computed following Ref.\ \cite{sokolov18} are also given for comparison.}
}
\begin{ruledtabular}
\begin{tabular}{lccccccc}
Configuration & Frequency difference $\Delta/\Gamma_0$ & Detuning $\delta/\Gamma_0$ & \red{$k_0 \ell_z$} & $k_0 \ell_z^*$ & $k_0 \ell_{\perp}^*$ & $h_z/\ell_z^*$ & $h_{\perp}/\ell_{\perp}^*$ \\
\colrule
~                                                          & 0  & 0.2 & \red{0.85} & \multicolumn{2}{c}{1.25} & \multicolumn{2}{c}{2.61}\\
~                                                          & 0  & 0.4 & \red{0.73} & \multicolumn{2}{c}{1.10} & \multicolumn{2}{c}{3.02}\\
$\vec{k}_{\mathrm{in}} \parallel \vec{E}_{\mathrm{ext}}$, $\vec{u}_{\mathrm{in}} = \;\circlearrowleft$   & 100  & 0.2 &  \red{0.6} & 1.12 & 1.88 & 1.91 & 1.06\\
$\vec{k}_{\mathrm{in}} \parallel \vec{E}_{\mathrm{ext}}$, $\vec{u}_{\mathrm{in}} = \;\circlearrowleft$   & 100  & 0.4 & \red{0.54} & 0.95 & 1.33 & 2.08 & 1.19\\
$\vec{k}_{\mathrm{in}} \perp \vec{E}_{\mathrm{ext}}$, $\vec{u}_{\mathrm{in}} = \;\uparrow$       & 100  & 0.2 & \red{0.44} & 0.71 & 1.24 & 1.36 & 1.14\\
$\vec{k}_{\mathrm{in}} \perp \vec{E}_{\mathrm{ext}}$, $\vec{u}_{\mathrm{in}} = \;\uparrow$      & 100  & 0.4 & \red{0.45} & 0.65 & 1.07 & 1.15 & 1.09\\
\end{tabular}
\end{ruledtabular}
\end{table*}

Symbols in Fig.\ \ref{figtrans}(b) show typical results for the average intensity transmission coefficient $\langle T(L, \omega) \rangle$ multiplied by the slab thickness $k_0 L$ in anticipation of the dependence $\langle T(L, \omega) \rangle \propto 1/L$ expected from the diffusion theory in the limit of $R, L \to \infty$ [see Eq.\ (\ref{transa}) in Appendix \ref{app:dif}]. The diffusion theory fits are shown in the same figure by solid lines. To obtain the fits we used Eq.\ (\ref{trans}) with $R_1 = R$, which accounts for the fact that light originating from the entire surface of the sample is collected when the intensity is measured at a large distance [$k_0 (z-L) = 10$ in Fig.\ \ref{figtrans}] behind the sample. The evolution of $\langle T(L, \omega) \rangle \times k_0 L$ with increasing $L$ turns out to be well captured by the diffusion theory that provides very good fits to the numerical data.
\red{This evolution is due to two reasons. First, the thicknesses of the slab $k_0 L \leq 10$ accessible for our numerical calculations would not be large enough compared to $k_0 \ell_z^* \simeq 1$ to ensure convergence of $\langle T(L, \omega) \rangle$ to a pure $1/L$ scaling even for a slab of infinite lateral extent $R \to \infty$. This yields $\langle T(L, \omega) \rangle \times k_0 L$ converging to a constant from below as $L$ increases and is sufficient to understand the two upper curves in Fig.\ \ref{figtrans}(b) corresponding to $\Delta = 0$. Second, the finite lateral size of the slab $k_0 R = 50$ speeds up the decrease of $\langle T(L, \omega) \rangle$ with $L$ because of the leakage of wave energy through the open lateral boundaries of the cylindrical sample. This effect starts to be visible for the four lower curves in Fig.\ \ref{figtrans}(b).} The best-fit parameters
\red{used for the theoretical curves in Fig.\ \ref{figtrans}(b)}
are summarized in Table\ \ref{tabfit}. Similarly to the fits in Figs.\ \ref{figlow} and \ref{figpop}, the fits in Fig.\ \ref{figtrans} may not be be unique and combinations of fit parameters different from those given in Table\ \ref{tabfit} may provide fits of comparable quality. This does not weaken our main conclusion about the diffuse nature of optical transport because the important aspect for us is the possibility of obtaining fits with reasonable fit parameters whereas the precise values of these parameters are not crucial for us in this work.

Analysis of best-fit parameters given in Table\ \ref{tabfit} allows us to make \red{several} important observations.
\red{First, the best-fit values of the transport mean free paths are consistently larger than the values of the scattering mean free path $\ell_z$ obtained by fitting the absolute value of the average atomic polarization to an exponentially decaying function $\exp(-z/2\ell_z)$ (see Ref.\ \cite{sokolov18} for details). This is a usual situation for multiple light scattering but no direct relation can be established between the scattering and transport mean free paths in a dense medium where the first-order perturbation theory in density $\rho$ is not expected to be valid. Second,}
values of $k_0 \ell_z^*$ as small as 0.65 are obtained, corresponding to $\ell_z^* \simeq 0.1 \lambda_0$. However, despite such a small value of the transport mean free path, the agreement of numerical results with the diffusion theory remains excellent. This is quite remarkable because a breakdown of diffusion could be expected for such a strong scattering based on the frequently used Ioffe-Regel criterion \cite{sheng95,skip18ir}.
\red{And finally}, the anisotropy of the transport mean free path deduced from the thickness dependence of the transmission coefficient is $\ell_{\perp}^*/\ell_z^* \sim 1.5$. The quantitative understanding of this anisotropy calls for development of analytic theory of light scattering in dense atomic media subjected to strong external electric fields, which is a formidable task falling beyond the scope of the present work.

\section{Conclusions}
\label{concl}

We performed numerical simulations of light transport through an optically thick, three-dimensional cloud of two-level atoms subjected to a strong static, external electric field. Both the average population of excited atomic states inside the cloud and the transmission coefficient of the cloud were calculated and analyzed for different frequencies and polarizations of the incident wave, and for different orientations of the external field. Comparison of numerical results with an analytic model of anisotropic photon diffusion indicates that the transport of optical energy in the atomic cloud can be perfectly described by the diffusion theory at least up to atomic number densities $\rho$ of the order of $10^2$ atoms per $\lambda_0^3$ (where $\lambda_0$ is the wavelength of light in the free space).  The electric field induces an optical anisotropy of the atomic medium, making the transport mean free paths vary by roughly 50\% depending on the spatial direction. At high atomic number densities $\rho$, the transport mean free path can become as small as $0.1 \lambda_0$. And still, diffusion holds and no signature of Anderson localization or any other mechanism of breakdown of diffusion, is found.
\red{It is quite remarkable that all our results are perfectly consistent with a constant, position-independent diffusion tensor $D$ with no need of introducing the position dependence of $D$ that might account for Anderson localization effects \cite{tiggelen00}.
}

Knowing that transport is diffusive is an important insight but the lacking theoretical element remains a full theoretical model for the transport lengths $\ell_z^*$ and $\ell_{\perp}^*$.
\red{Even for dilute atomic media we didn't find any results for $\ell_z^*$ and $\ell_{\perp}^*$ in the presence of an external electric field in the literature despite the fact that the expression for the atomic polarizability is simple \cite{sokolov17,larionov18}. For high densities,} the problem is difficult to solve even in the absence of external fields when $\ell^*$ has been calculated only up to second order in $\rho/k_0^3$ \cite{tiggelen94,cherroret16}. Numerical methods employed in this work proved to be very useful to guide and test analytical theories in this research field \cite{kwong18}.

\begin{acknowledgments}
This work was funded by the Agence Nationale de la Recherche (Grant No. ANR-14-CE26-0032 LOVE).
Quantum-mechanical calculations of the spatial distributions of excited state populations and of the transmission coefficients of atomic clouds were carried out with the financial support of the Russian Science Foundation (Project No. 17-12-01085). IMS acknowledges the hospitality of the LPMMC where a part of this work has been performed with a financial support of the Centre de Physique Th\'{e}orique de Grenoble-Alpes (CPTGA).

\end{acknowledgments}

\appendix
\section{Quantum-mechanical model of light scattering by two-level atoms in a static electric field }
\label{app:qm}

$N$ immobile two-level atoms (ground state $\left| {\cal E}_g, J_g = 0 \right>$, excited states $\left| {\cal E}_e, J_e = 1, m = 0, \pm 1 \right>$) in a static and spatially uniform electric field, interacting via the free electromagnetic field, can be described by the following approximate Hamiltonian \cite{sokolov17,sokolov18,skip19}:
\begin{eqnarray}
{\hat H} &=& \sum\limits_{j=1}^{N} \sum\limits_{m=-1}^{1} \hbar \left( \omega_0 - m^2 \Delta \right)
\nonumber \\
&\times& \left| {\cal E}'_e(m), J_e = 1, m \right>_j \left< {\cal E}'_e(m), J_e = 1, m \right|_j
\nonumber \\
&+&
\sum\limits_{\bm{\epsilon} \perp \mathbf{k}} \hbar ck
\left( {\hat a}_{\mathbf{k} \bm{\epsilon}}^{\dagger} {\hat a}_{\mathbf{k}\bm{\epsilon}} + \frac12 \right)
- \sum\limits_{j=1}^{N} {\hat{\mathbf{D}}}_j \cdot {\hat{\mathbf{E}}}(\mathbf{r}_j)
\nonumber \\
&+& \frac{1}{2 \varepsilon_0}
\sum\limits_{j \ne n}^{N} {\hat{\mathbf{D}}}_j \cdot {\hat{\mathbf{D}}}_n \delta(\mathbf{r}_j - \mathbf{r}_n),
\label{ham}
\end{eqnarray}
where ${\cal E}_{e}'(m)$ are the energies of excited states having a magnetic quantum number $m_e = m$ in the electric field,
$\omega_0$ is the frequency of the transition $\left| {\cal E}_g', J_g = 0 \right> \to \left| {\cal E}'_e(m), J_e = 1, m=0 \right>$, $\hbar \Delta$ is the energy difference between the excited states with $m = 0$ and $m = \pm 1$ due to Stark shifts, ${\hat a}_{\mathbf{k} \bm{\epsilon}}^{\dagger}$ and ${\hat a}_{\mathbf{k}\bm{\epsilon}}$ are creation and annihilation operators corresponding to an electromagnetic mode with a wave vector $\mathbf{k}$ and a polarization $\bm{\epsilon}$,
${\hat{\mathbf{D}}}_j$ are atomic dipole operators,
$\varepsilon_0 {\hat{\mathbf{E}}}(\mathbf{r}_j)$ are electric displacement vectors at atomic positions $\vec{r}_j$,
\red{and the quantization axis is chosen parallel to the external electric field.}

The Hamiltonian (\ref{ham}) is quite general and can be used to describe many physical phenomena arising from the interaction of light with atoms, including nonlinear effects. Here we restrict our consideration to the linear optics regime which, strictly speaking, corresponds to allowing only a single excitation (photon) in the system. In reality our results will apply when the number of excitations is much less than the number of atoms, which implies low intensity of incident light in an experiment. For a single excitation, Eq.\ (\ref{ham}) reduces to an effective non-Hermitian Hamiltonian of the atomic subsystem \cite{sokolov17,sokolov18,skip19}:
\begin{eqnarray}
G_{e_{j m} e_{n m'}} &=& \left(i + 2 m^2 \frac{\Delta}{\Gamma_0} \right) \delta_{e_{j m} e_{n m'}} +
\frac{2 k_0^3}{\hbar \Gamma_0} (1 - \delta_{e_{j m} e_{n m'}})
\nonumber \\
&\times&
\sum\limits_{\mu, \nu}
{d}_{e_{j m} g_j}^{\mu} {d}_{g_n e_{n m'}}^{\nu}
\frac{e^{i k_0 r_{jn}}}{k_0 r_{jn}}
\nonumber
\\
&\times& \left[
\vphantom{\frac{r_{jn}^{\mu} r_{jn}^{\nu}}{r_{jn}^2}}
 \delta_{\mu \nu}
P(i k_0 r_{jn})
+ \frac{r_{jn}^{\mu} r_{jn}^{\nu}}{r_{jn}^2}
Q(i k_0 r_{jn})
\right],
\label{green}
\end{eqnarray}
where $P(x) = 1 - 1/x + 1/x^2$, $Q(x) = -1 + 3/x - 3/x^2$,
$\Gamma_0$ is the natural line width of the excited states of an isolated atom,
$\vec{d}_{e_{j m} g_j} = \langle {\cal E}_e'(m), J_{e}=1, m|{\hat{\mathbf{D}}}_j | {\cal E}_g', J_{g} = 0 \rangle$, and $\vec{r}_{jn} = \vec{r}_j - \vec{r}_n$. The $3N \times 3N$ matrix (\ref{green}) describes the system of $N$ atoms coupled via electromagnetic fields. Its non-Hermiticity is due to the openness of the atomic system and the leakage of energy out of it via emission of light (photons). Properties of eigenvalues and eigenvectors of matrix $G$ have been studied in Ref.\ \cite{skip19}.

\section{Solution of the anisotropic diffusion equation for light in a disordered medium}
\label{app:dif}

In this Appendix, we present a solution of the anisotropic diffusion equation (\ref{difeq}) with the boundary conditions (\ref{bc1}) and (\ref{bc2}) in a cylindrical sample depicted in the inset of Fig.\ \ref{figlow}. Equation (\ref{difeq}) can be recast as an isotropic diffusion equation
\begin{eqnarray}
-\bm{\nabla_{\vec{r}'}}^2 \red{P_{\mathrm{dif}}}(\vec{r}') = \frac{3 \red{P_0}}{\ell_z^*} \delta(z'-\ell_{z}^*)
\Pi\left(\frac{r_{\perp}'}{2R'} \right),
\label{difeq2}
\end{eqnarray}
where $\vec{r}' = \{ \vec{r}_{\perp}' = \vec{r}_{\perp} \ell_z^*/\ell_{\perp}^*, z' = z \}$. The boundary conditions (\ref{bc1}) and (\ref{bc2}) preserve their form with $R$ and $h_{\perp}$ replaced by $R' = R \ell_z^*/\ell_{\perp}^*$ and  $h_{\perp}' = h_{\perp} \ell_z^*/\ell_{\perp}^*$, respectively.

A solution of Eq.\ (\ref{difeq2}) that remains finite for $r_{\perp}' \to 0$ can be represented as
\begin{eqnarray}
\red{P_{\mathrm{dif}}}(\vec{r}') &=& \sum\limits_{n=1}^{\infty} J_0(\kappa_n r_{\perp}')
\nonumber \\
&\times& \left[
A_n \sinh \kappa_n z + B_n \cosh \kappa_n z
\right],
\label{sol1}
\end{eqnarray}
where $J_0$ is the zeroth-order Bessel junction.
The coefficients $\kappa_n$ are found from the boundary condition (\ref{bc2}): $\kappa_n = \beta_n/(R' + h_{\perp}')$, where $\beta_n$ denotes the $n$-th zero of the Bessel junction $J_0$: $J_0(\beta_n) = 0$. The coefficients $A_n$ and $B_n$ follow from the boundary condition (\ref{bc1}) and the explicit form of the source term in Eq.\ (\ref{difeq2}) with an identity
\begin{eqnarray}
\Pi\left(\frac{r_{\perp}'}{2R'} \right) =
\frac{2 R'}{R' + h_{\perp}'} \sum\limits_{n=1}^{\infty}
J_0(\kappa_n r_{\perp}') \frac{J_1(\kappa_n R')}{J_1(\beta_n)^2 \beta_n},\;\;
\label{piidentity}
\end{eqnarray}
where $J_1$ denotes the first-order Bessel function.

We finally obtain
\begin{eqnarray}
&& \red{P_{\mathrm{dif}}}(\vec{r}') = 6 \red{P_0} \frac{R'}{R' + h_{\perp}'} \sum\limits_{n=1}^{\infty} J_0(\kappa_n r_{\perp}')
\frac{J_1(\kappa_n R')}{J_1(\beta_n)^2 \beta_n}
\nonumber \\
&&\;\; \times
\frac{\sinh[\kappa_n(z_< + h_z)] \sinh[\kappa_n(L + h_z - z_>)]}{\kappa_n \ell_z^* \sinh[\kappa_n(L + 2 h_z)]},
\label{sol2}
\end{eqnarray}
where $z_{<} = \min(z, \ell_z^*)$ and $z_{>} = \max(z, \ell_z^*)$.
For an infinitely wide slab $R' \to \infty$, Eq.\ (\ref{sol2}) reduces to
\begin{eqnarray}
\red{P_{\mathrm{dif}}}(\vec{r}') = \frac{3 \red{P_0}}{\ell_z^*} \times \frac{(z_< + h_z)(L + h_z - z_>)}{L + 2 h_z},
\label{sol2a}
\end{eqnarray}
where we used the fact that
\begin{eqnarray}
\sum\limits_{n=1}^{\infty} \frac{1}{J_1(\beta_n) \beta_n} = \frac{1}{2}.
\label{sum}
\end{eqnarray}
We see from Eq.\ (\ref{sol2a}) that the
\red{solution}
is not sensitive to the anisotropy of the scattering medium in an infinitely wide slab illuminated by a plane wave.

In the main text, we analyze the $z$-dependence of
\red{average population of excited states} averaged over a circular area of radius $R_1 < R$ in the central part of the cylindrical sample. This quantity readily follows from Eq.\ (\ref{sol2}):
\begin{eqnarray}
&& \langle \red{P_{\mathrm{dif}}}(\vec{r}') \rangle_{R_1'} =
\frac{1}{\pi R_1'^2} \int\limits_{r_{\perp}' < R_1'} \red{P_{\mathrm{dif}}}(\vec{r}') d^2 r_{\perp}'
\nonumber \\
&&\;\; = 6 \red{P_0} \frac{R'}{R' + h_{\perp}'} \sum\limits_{n=1}^{\infty}
\frac{2 J_1(\kappa_n R_1')}{\kappa_n R_1'}
\frac{J_1(\kappa_n R')}{J_1(\beta_n)^2 \beta_n}
\nonumber \\
&&\;\; \times
\frac{\sinh[\kappa_n(z_< + h_z)] \sinh[\kappa_n(L + h_z - z_>)]}{\kappa_n \ell_z^* \sinh[\kappa_n(L + 2 h_z)]}.
\label{sol3}
\end{eqnarray}

The average intensity transmission coefficient is
\red{obtained by noting that in the diffusion approximation, the average intensity in the atomic medium is expected to obey the same diffusion equations (\ref{difeq}) and (\ref{difeq2}) as $\red{P_{\mathrm{dif}}}$:}
\begin{eqnarray}
\langle T(L) \rangle &=& -\frac{1}{\red{P_0}}\, \frac{\ell^*}{3}\, \left. \frac{\partial}{\partial z}
\langle \red{P_{\mathrm{dif}}}(\vec{r}') \rangle_{R_1'} \right|_{z = L}
\nonumber \\
&=&
\frac{2 R'}{R' + h_{\perp}'} \sum\limits_{n=1}^{\infty}
\frac{2 J_1(\kappa_n R_1')}{\kappa_n R_1'}
\frac{J_1(\kappa_n R')}{J_1(\beta_n)^2 \beta_n}
\nonumber \\
&&\;\; \times
\frac{\sinh[\kappa_n(\ell_z^* + h_z)] \cosh[\kappa_n h_z]}{\sinh[\kappa_n(L + 2 h_z)]}.
\label{trans}
\end{eqnarray}
For $R' \to \infty$, this expression reduces to
\begin{eqnarray}
\langle T(L) \rangle &=& \frac{\ell_z^* + h_z}{L + 2 h_z}.
\label{transa}
\end{eqnarray}
Similarly to \red{Eq.}\ (\ref{sol2a}), this result is not sensitive to the anisotropy of the medium.


\begin{thebibliography}{99}

\bibitem{lagen96}
A. Lagendijk and B.A. van Tiggelen,
Resonant multiple scattering of light,
Phys. Rep. \textbf{270}, 143 (1996).

\bibitem{devries98}
P. de Vries, D.V. van Coevorden, and A. Lagendijk,
Point scatterers for classical waves,
Rev. Mod. Phys. \textbf{70}, 447 (1998).

\bibitem{phillips98}
W.D. Phillips,
Nobel Lecture: Laser cooling and trapping of neutral atoms,
Rev. Mod. Phys. \textbf{70}, 721 (1998).

\bibitem{nieuwen94}
Th.M. Nieuwenhuizen, A.L. Burin, Yu. Kagan, and G.V. Shlyapnikov,
Light propagation in a solid with resonant atoms at random positions,
Phys. Lett. A \textbf{184}, 360 
(1994).

\bibitem{tiggelen94}
B.A. van Tiggelen and A. Lagendijk,
Resonantly induced dipole-dipole interactions in the diffusion of scalar waves,
Phys. Rev. B \textbf{50}, 16729 (1994).

\bibitem{cherroret16}
N. Cherroret, D. Delande, and B.A. van Tiggelen,
Induced dipole-dipole interactions in light diffusion from point dipoles,
Phys. Rev. A \textbf{94}, 012702 (2016).

\bibitem{jenkins16}
S.D. Jenkins, J. Ruostekoski, J. Javanainen, R. Bourgain, S. Jennewein, Y.R.P. Sortais, and A. Browaeys,
Optical Resonance Shifts in the Fluorescence of Thermal and Cold Atomic Gases,
Phys. Rev. Lett. \textbf{116}, 183601 (2016).

\bibitem{kwong18}
C.C. Kwong, D. Wilkowski, D. Delande, and R. Pierrat,
Coherent light propagation through cold atomic clouds beyond the independent scattering approximation,
Phys. Rev. A \textbf{99}, 043806 (2019).

\bibitem{fayard15}
N. Fayard, A. Caz\'{e}, R. Pierrat, and R. Carminati,
Intensity correlations between reflected and transmitted speckle patterns,
Phys. Rev. A \textbf{92}, 033827 (2015).

\bibitem{cottier18}
F. Cottier, A. Cipris, R. Bachelard, and R. Kaiser,
Intensity fluctuations signature of 3D Anderson localization of light,
arXiv:1812.10313.

\bibitem{skip14}
S.E. Skipetrov and I.M. Sokolov,
Absence of Anderson localization of light in a random ensemble of point scatterers,
Phys. Rev. Lett. \textbf{112}, 023905 (2014).

\bibitem{bellando14}
L. Bellando, A. Gero, E. Akkermans, and R. Kaiser,
Cooperative effects and disorder: A scaling analysis of the spectrum of the effective atomic Hamiltonian,
Phys. Rev. A \textbf{90}, 063822 (2014).

\bibitem{skip15}
S.E. Skipetrov and I.M. Sokolov,
Magnetic-Field-Driven Localization of Light in a Cold-Atom Gas,
Phys. Rev. Lett. \textbf{114}, 053902 (2015).

\bibitem{anderson58}
P.W. Anderson,
Absence of diffusion in certain random lattices,
Phys. Rev. \textbf{109}, 1492 
(1958).

\bibitem{lagendijk09}
A. Lagendijk, B. A. van Tiggelen, and D. S. Wiersma,
Fifty years of Anderson localization,
Phys. Today \textbf{62}(8), 24 (2009).

\bibitem{abrahams10}
E. Abrahams (ed.),
\textit{50 Years of Anderson Localization}
(World Scientific, Singapore, 2010).

\bibitem{kramer93}
B. Kramer and A. MacKinnon,
Localization: theory and experiment,
Rep. Prog. Phys. \textbf{56}, 1469 
(1993).

\bibitem{chabe08}
J. Chab\'{e}, G. Lemari\'{e}, B. Gr\'{e}maud, D. Delande, P. Szriftgiser, and J.C. Garreau,
Experimental observation of the Anderson metal-insulator transition with atomic matter waves,
Phys. Rev. Lett. \textbf{101}, 255702 (2008).

\bibitem{jendr12}
F. Jendrzejewski, A. Bernard, K. M\"{u}ller, P. Cheinet, V. Josse, M. Piraud, L. Pezz\'{e}, L. Sanchez-Palencia, A. Aspect, and P. Bouyer,
Three-dimensional localization of ultracold atoms in an optical disordered potential,
Nature Phys. \textbf{8}, 398 
(2012).

\bibitem{hu08}
H. Hu, A. Strybulevych, J.H. Page, S.E. Skipetrov, and B.A. van Tiggelen,
Localization of ultrasound in a three-dimensional elastic network,
Nat. Phys. \textbf{4}, 945 
(2008).

\bibitem{cobus18}
L.A. Cobus, W.K. Hildebrand, S.E. Skipetrov, B.A. van Tiggelen, and J.H. Page,
Transverse confinement of ultrasound through the Anderson transition in three-dimensional mesoglasses,
Phys. Rev. B \textbf{98}, 214201 (2018).

\bibitem{chabanov00}
A.A. Chabanov, M. Stoytchev, and A.Z. Genack,
Statistical signatures of photon localization,
Nature \textbf{404}, 850 
(2000).

\bibitem{schwartz07}
T. Schwartz, G. Bartal, S. Fishman, and M. Segev,
Transport and Anderson localization in disordered two-dimensional photonic lattices,
Nature \textbf{446}, 52 
(2007).

\bibitem{skip16}
S.E. Skipetrov and J.H. Page,
Red light for Anderson localization,
New J. Phys. \textbf{18}, 021001 (2016).

\bibitem{skip16pra}
S.E. Skipetrov, I.M. Sokolov, and M.D. Havey,
Control of light trapping in a large atomic system by a static magnetic field,
Phys. Rev. A \textbf{94}, 013825 (2016).

\bibitem{skip19}
S.E. Skipetrov and I.M. Sokolov,
Search for Anderson localization of light by cold atoms in a static electric field,
Phys. Rev. B \textbf{99}, 134201 (2019).

\bibitem{araujo17}
\red{
M.O. Ara\'{u}jo, W. Guerin, and R. Kaiser,
Decay dynamics in the coupled-dipole model,
J. Mod. Opt. \textbf{65}, 1345 (2017).
}


\bibitem{friedrich90}
H. Friedrich,
\textit{Theoretical Atomic Physics} (Springer, Berlin, 1990).

\bibitem{sobelman92}
I.I. Sobelman,
\textit{Atomic Spectra and Radiative Transitions. 2nd Edition} (Springer, Berlin, 1992).

\bibitem{fofanov13}
Ya.A. Fofanov, A.S. Kuraptsev, I.M. Sokolov, and M.D. Havey,
Spatial distribution of optically induced atomic excitation in a dense and cold atomic ensemble,
Phys. Rev. A \textbf{87}, 063839 (2013).

\bibitem{lax51}
\red{M. Lax,
Multiple scattering of waves,
Rev. Mod. Phys. \textbf{23}, 287 (1951).
}

\bibitem{ishimaru78}
A. Ishimaru,
\textit{Wave Propagation and Scattering in Random Media}
(Academic, New York, 1978).

\bibitem{sheng95}
P. Sheng,
\textit{Introduction to Wave Scattering, Localization and Mesoscopic Phenomena}
(Academic Press, San Diego, 1995).

\bibitem{akkermans07}
E. Akkermans and G. Montambaux,
\textit{Mesoscopic Physics of Electrons and Photons}
(Cambridge University Press, Cambridge, 2007).

\bibitem{stark96}
H. Stark  and  T.C.Lubensky,
Multiple Light Scattering in Nematic Liquid Crystals,
Phys. Rev. Lett. \textbf{77}, 2229 (1996).

\bibitem{tiggelen96}
B.A. van Tiggelen, R. Maynard, and A. Heiderich,
Anisotropic Light Diffusion in Oriented Nematic Liquid Crystals,
Phys. Rev. Lett. \textbf{77}, 639 (1996).

\bibitem{kao96}
M.H. Kao, K.A. Jester, A.G. Yodh, and P.J. Collings,
Observation of Light Diffusion and Correlation Transport in Nematic Liquid Crystals,
Phys. Rev. Lett. \textbf{77}, 2233 (1996).

\bibitem{wiersma99}
D.S. Wiersma, A. Muzzi, M. Colocci, and R. Righini,
Time-Resolved Anisotropic Multiple Light Scattering in Nematic Liquid Crystals,
Phys. Rev. Lett. \textbf{83}, 4321 (1999).

\bibitem{johnson08}
P.M. Johnson, S. Faez, and A. Lagendijk,
Full characterization of anisotropic diffuse light,
Opt. Express \textbf{16}, 7435 (2008).

\bibitem{labeyrie03}
G. Labeyrie, E. Vaujour, C. A. M\"{u}ller, D. Delande, C. Miniatura, D. Wilkowski, and R. Kaiser,
Slow Diffusion of Light in a Cold Atomic Cloud,
Phys. Rev. Lett. \textbf{91}, 223904 (2003).

\bibitem{zhu91}
J.X. Zhu, D.J. Pine, and D.A. Weitz,
Internal reflection of diffusive light in random media,
Phys. Rev. A \textbf{44}, 3948 (1991).

\bibitem{morice95}
\red{
O. Morice, Y. Castin, and J. Dalibard,
Refractive index of a dilute Bose gas,
Phys. Rev. A \textbf{51}, 3896 (1995).
}

\bibitem{java17}
\red{
J. Javanainen, J. Ruostekoski, Y. Li, and S.-M. Yoo,
Exact electrodynamics versus standard optics for a slab of cold dense gas,
Phys. Rev. A \textbf{96}, 033835 (2017).
}

\bibitem{skip18ir}
S.E. Skipetrov and I.M. Sokolov,
Ioffe-Regel criterion of Anderson localization in the model of resonant point scatterers,
Phys. Rev. B \textbf{98}, 064207 (2018).

\bibitem{sokolov17}
I.M. Sokolov,
Influence of an electrostatic field on the permittivity of dense and cold atomic ensembles,
JETP Lett. \textbf{106}, 341 (2017).

\bibitem{sokolov18}
I.M. Sokolov,
Electro-optical effects in dense and cold atomic gases,
Phys. Rev. A \textbf{98}, 013412 (2018).

\bibitem{tiggelen00}
\red{
B.A. van Tiggelen, A. Lagendijk, and D.S. Wiersma,
Reflection and Transmission of Waves near the Localization Threshold,
Phys. Rev. Lett. \textbf{84}, 4333 (2000).
}

\bibitem{larionov18}
\red{
N.V. Larionov and I.M. Sokolov,
Influence of electric and magnetic fields on interference effects upon multiple light scattering in cold atomic ensembles,
J. Exp. Theor. Phys. \textbf{127}, 264 (2018).
}


\end{thebibliography}
\end{document}